\documentclass[manuscript, acmsmall]{acmart}

\usepackage{amsmath}
\usepackage{mathtools}
\usepackage{fixltx2e}
\usepackage{graphicx}
\usepackage{relsize}
\usepackage{todonotes}
\usepackage{array, subcaption, multirow}
\usepackage{caption}

\newcommand\floor[1]{\lfloor#1\rfloor}

\AtBeginDocument{%
  \providecommand\BibTeX{{%
    \normalfont B\kern-0.5em{\scshape i\kern-0.25em b}\kern-0.8em\TeX}}}

\setcopyright{acmlicensed}
\copyrightyear{2024}
\acmYear{2024}
\acmDOI{10.1145/0000000.0000000}

\acmConference[CSCW '24]{Proceedings of the 27th ACM Conference on Computer-Supported Cooperative Work and Social Computing}{Nov 9 -- 13, 2024}{San Jose, Costa Rica}
\acmBooktitle{Proceedings of the 27th ACM Conference on Computer-Supported Cooperative Work and Social Computing (CSCW '24), Nov 9 -- 13, 2024, San Jose, Costa Rica}
\acmPrice{15.00}
\acmISBN{000-0-0000-0000-0/00/00}

\begin{document}

\title[Anon.]{Simplify, Consolidate, Intervene: Facilitating Institutional Support with Mental Models of Learning Management System Use}

\author{Taha Hassan}
\affiliation{%
  \institution{Department of Computer Science, Virginia Tech}
  \city{Blacksburg}
  \state{Virginia}
\country{United States}}
\email{taha@vt.edu}

\author{Bob Edmison}
\affiliation{%
  \institution{Department of Computer Science, Virginia Tech}
  \city{Blacksburg}
  \state{Virginia}
\country{United States}}
\email{kedmison@vt.edu}

\author{Daron Williams}
\affiliation{%
  \institution{Technology-Enhanced Learning and Online Strategies, Virginia Tech}
  \city{Blacksburg}
  \state{Virginia}
\country{United States}}
\email{debo9@vt.edu}

\author{Larry Cox II}
\affiliation{%
  \institution{Technology-Enhanced Learning and Online Strategies, Virginia Tech}
  \city{Blacksburg}
  \state{Virginia}
\country{United States}}
\email{lacox@vt.edu}

\author{Matthew Louvet}
\affiliation{%
  \institution{Technology-Enhanced Learning and Online Strategies, Virginia Tech}
  \city{Blacksburg}
  \state{Virginia}
\country{United States}}
\email{mattl06@vt.edu}

\author{Bart Knijnenburg}
\affiliation{%
  \institution{School of Computing, Clemson University}
  \city{Clemson}
  \state{South Carolina}
\country{United States}}
\email{bartk@clemson.edu}

\author{D. Scott McCrickard}
\affiliation{%
  \institution{Department of Computer Science, Virginia Tech}
  \city{Blacksburg}
  \state{Virginia}
\country{United States}}
\email{mccricks@vt.edu}
\renewcommand{\shortauthors}{Hassan, et al.}

\begin{abstract}
Measuring instructors' adoption of learning management system (LMS) tools is a critical first step in evaluating the efficacy of online teaching and learning at scale. Existing models for LMS adoption are often qualitative, learner-centered, and difficult to leverage towards institutional support. We propose depth-of-use (DOU): an intuitive measurement model for faculty's utilization of a university-wide LMS and their needs for institutional support. We hypothesis-test the relationship between DOU and course attributes like modality, participation, logistics, and outcomes. In a large-scale analysis of metadata from 30000+ courses offered at Virginia Tech over two years, we find that a pervasive need for scale, interoperability and ubiquitous access drives LMS adoption by university instructors. We then demonstrate how DOU can help faculty members identify the opportunity-cost of transition from legacy apps to LMS tools. We also describe how DOU can help instructional designers and IT organizational leadership evaluate the impact of their support allocation, faculty development and LMS evangelism initiatives. 
\end{abstract}

\keywords{learning management system, adoption, context, institutional support, assessment, instructional design, higher education}

\maketitle

\section{Introduction}
Learning management systems (LMS) are increasingly the primary infrastructure for hosting and disseminating information between key stakeholders in the higher education domain \citep{coates2005critical,lmsadoption}. A contemporary service-based (SaaS) LMS is, typically, a compendium of online communication, productivity, assessment, and class-management applications. In recent years, LMSs have received widespread adoption across the global educational IT landscape \citep{mattes2017private, marachi2020case, saroia2019investigating,garone2019clustering, sezer2019learning,isaac2019online, bervell2020blended}. They support a diverse array of teaching and learning practices, including remote teaching, self-directed learning, mobile learning, and computer-supported collaborative work. UNESCO's global project to address school closures during the COVID-19 global pandemic \citep{unesco} lists digital LMSs among the tools recommended in ensuring teacher-learner connectivity, supporting open learning, and fostering community participation in teaching. Understanding the adoption and impact of LMS apps and services is, therefore, central to faculty, university administrators, and instructional designers in better designing and evaluating course content. There has been ample work on qualitative driving factors of LMS adoption \citep{berggren2005practical,west2007understanding,webct,mtebe2015learning, khechine2020adoption}. However, this research is largely limited to self-reported LMS use, and does not account for individual LMS tools or stakeholder priorities. Furthermore, there is no analytic consensus on how to model frequent LMS use-contexts, and test their relationship with learning outcomes, especially at scale. Note that a \textit{use-context} is any meaningful set of course attributes with a potential impact on learning outcomes. This includes myriad aspects of course content, mode-of-delivery, participation, and logistics. This multitude of conditions, a variety of LMS data sources (app metadata, course site content, team drives, social media), and the large volume of raw LMS page requests, pose significant challenges to data aggregation, reporting, and supporting institutional policy-making. Martin et al.'s summary of review studies on online education research spanning three decades (1990-2018) \cite{martin2020systematic} identifies three broad research domains: "course+instructor" (\textbf{CI}), "organization" (\textbf{ORG}), and "learner" (\textbf{L}). The review authors observe that themes within the "learner" domain, especially learner engagement and learner profiling have received considerable attention by the community. They simultaneously identify the need for additional research on the less-frequently studied \textbf{CI} and \textbf{ORG} themes. These include course technologies (LMS, wikis, web conferencing, social networking), institutional support (faculty mentoring, professional development), and departmental policy-making (resource allocation to stakeholders, managing online teaching, inclusivity, and ethics).

\begin{figure*}
\centering
\includegraphics[width=0.8\textwidth]{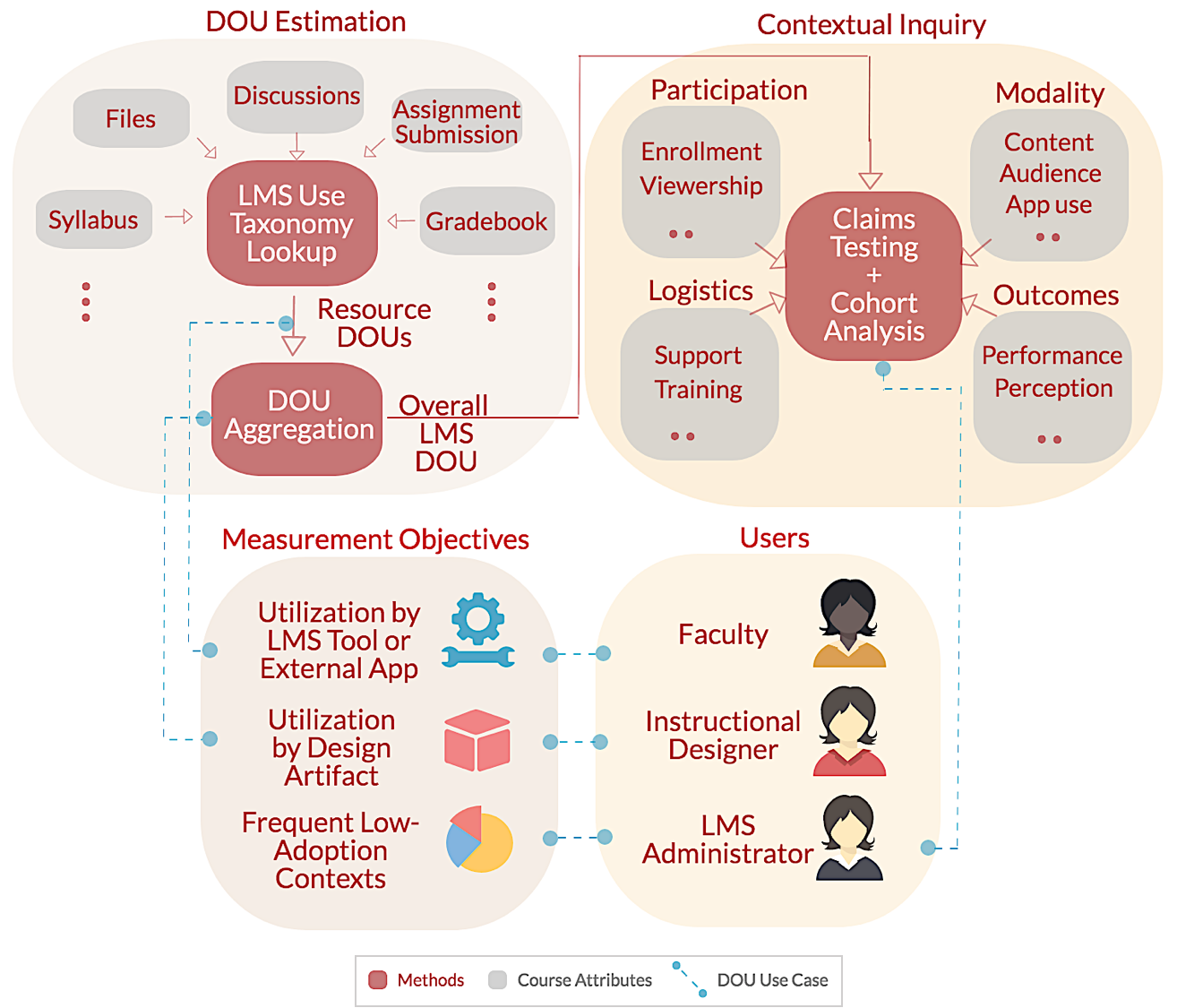}
\caption{Study methodology overview (clockwise from top left): DOU Estimation (data sources and methods), Contextual Inquiry (hypothesis variables and methods), key user groups, and measurement objectives fulfilled by DOU. (Top-left) We estimate a DOU score of overall LMS use - low, medium or high - for each course in our analysis. We then test claims (top-right) about how DOU is linked to course modality, participation, logistics, and outcomes. Finally, we discuss DOU use-cases (bottom-left and bottom-right) for faculty, instructional designers, and LMS administrators.}
\label{fig_sys_overview}
\end{figure*}

These research domains thus inform the two fundamental research questions in our study. (\textbf{RQ1}, \textbf{CI}) What aspects of the course design, content and delivery drive faculty to \textit{adopt} or ignore individual LMS tools? (\textbf{RQ2}, \textbf{ORG}) How can we produce quantitative, reproducible insights about faculty's LMS use to inform \textit{institutional support and management}?

To answer these research questions, we develop a novel model for measurement of overall LMS use by course faculty, called the `Depth of Use' (DOU). Using this model, we assign an ordinal DOU score (low, medium or high) to each of over thirty-thousand college courses offered between, and including, spring 2021 and spring 2023 at Virginia Tech. We then test ten hypotheses about how DOU is linked to course attributes like modality, participation, logistics and outcomes. This allows us to determine the conditions in which new LMS tools are adopted or ignored (\textbf{RQ1}), and course cohorts respond to faculty support and professional development (\textbf{RQ2}). For instance, we discover that faculty rely on tools that favor scale, ubiquitous access and interoperability, and overall LMS use is linked to better learning outcomes. We also provide case studies of several DOU applications we developed at Virginia Tech. These include cost-benefit analysis for adoption of new LMS tools, supporting course evaluation and LMS evangelism through cohort analysis, and resource allocation for remote teaching during COVID-19. Figure \ref{fig_sys_overview} describes our overall approach, and section 3 details the DOU measurement model. Table \ref{tab_taxonomy} describes the vendor-agnostic taxonomy of LMS use informing our DOU model. Overall, our framework forwards a novel multistakeholder view of LMS utilization, in that alongside learning analytics, it supports claim testing and cohort analysis for policy decision-making, an avenue with lesser treatment in prior literature on online teaching and learning in the last decade \citep{martin2020systematic}.

We make the following contributions in our study:

\begin{enumerate}
    \item We present DOU, an intuitive, actionable model of learning management system use, leveraging the experiences of Virginia Tech instructional designers leading a multiyear course redesign initiative (Section 3),

    \item We evaluate the relationship between learning management system use and course attributes like modality, participation, logistics and outcomes for a two year period (Section 4),

    \item We identify two key application areas of DOU: adoption contexts and institutional support, and describe key supporting analyses to enable decision-making in the management of higher learning (Section 5), 
    
\end{enumerate}

The rest of this paper is organized as follows. Section 2 extends our analysis of related work. Section 3 describes the depth-of-use (DOU) measurement model, and the research questions and hypotheses examined in our study. Section 4 details the datasets, methods and results from hypothesis tests performed on DOU and its constituent dimensions. Section 5 and 6 conclude the study with a discussion of DOU use-cases for faculty, instructional designers, and LMS administrators. 

\section{Related work}
\subsection{LMS adoption: human factors and information systems}
There is considerable prior work on qualitative grounds for LMS adoption, like teaching and learning efficiency, generational student expectations, and institutional expansion and consolidation \citep{coates2005critical,west2007understanding}. For course instructors, the basic predictors of the pace of LMS adoption are departmental affiliation (STEM vs. non-STEM, say) and course modality (online vs. face-to-face, say). \citet{west2007understanding} conducted semi-structured interviews with 30 college instructors over two semesters, about primary use cases, teaching efficacy and efficiency, and overall satisfaction with Blackboard LMS. The study identified so-called `integration challenges': course instructors finding it difficult to integrate LMS services into their teaching practices. This notion of `integration' was echoed by \citet{mcgill2009task} for the case of student adoption of WebCT, whereby students with a more favorable view of the `task-technology fit' of LMS services were more likely to have higher LMS utilization. The authors also noted that instructor norms (instructor's view of LMS usability, support staff availability, and access to training resources) affected student utilization of LMS services favorably. Following an institution-wide transition to Canvas LMS, \citet{wilcox2016canvas} surveyed user perceptions on frequent modes of use and platform limitations for Canvas LMS. They identified a generation gap in expectations between students and course instructors, wherein the pervasive student use of the mobile LMS app rendered a subset of Canvas sites - designed by faculty members for the desktop - ineffective in navigation, flow and content organization.

Likewise, an information systems (IS) perspective on LMS adoption has been thoroughly explored over the years \citep{webct,blackboard,ozkan}. A bulk of these studies apply and evaluate a canonical model of IS success first discussed by \citet{delone1992}. The model factorizes the individual and organizational success of an IS into quality (system, information, and service), use (utilization, intention of use) and net benefits (impact on overall satisfaction, and intention of use) \citep{delone2003}. \citet{webct} conducted a university-wide study of IS success factors underlying WebCT adoption and operationalized LMS utilization using nature of use (mandatory or optional), frequency of use, access and availability. They found use and intention of use both to be strong correlates of WebCT success. \citet{fathema2015expanding} evaluated TAM using survey data on faculty and student attitudes about Canvas LMS at two public universities. They discovered that system quality and user self-efficacy were strongly linked to system use and perceived usefulness. They also noted that system quality is a multi-faceted notion that incorporates issues like design aesthetics, flexibility of access, degree of customization, and multimedia support. \citet{ngai2007empirical} reported a stronger effect of the perceived usefulness and ease-of-use on system use relative to that of attitude (interest expressed towards adopting a new system). These studies largely employ user-reported system use in their analyses. Nonetheless, there are some early instances of LMS use modeling such as \citet{ozkan}, where study participants reported system use as the number of hours spent daily, on course-related activities with U-Link using a desktop or web application.

\subsection{Institutional support and management}
Review studies by \cite{tallent2006teaching}, \cite{zawacki2009review}, and \cite{martin2020systematic} note that an instructive, albeit limited body of educational research exists on institutional and department-level inquiries into faculty mentoring and professional development \citep{manduca2017improving}, technical support \citep{sitzmann2010effects}, instructional consultations \citep{finelli2008utilizing}, online teaching policies \citep{jones2008issues, martin2017global}, faculty incentive schemes \citep{rockwell1999incentives}, and support systems for students \citep{moisey2008supporting}. The COVID-19 pandemic has brought renewed attention, for instance, to faculty needs that hinder the development of online and distance education coursework. In a case study of obstacles to distance education \citep{rockwell1999incentives}, faculty and administrators cited the need for instructional design and technical support to aid digital skills development, and reduce the time overhead of course development. In \cite{sitzmann2010effects}, the authors found that technical difficulties in a digital skills training lowered the participants' test scores, and challenged their pre-training motivation. They recommended organizations invest in technical support, outreach to motivate potential learners, and interruption-free training environments. In comparison, a study by \cite{jones2008issues} examined frequent concerns voiced by administrators in selecting online technologies for post-secondary distance learning. Cost (time, money and manpower) of delivery and support, especially at scale, was found to be the biggest concern, followed by vendor lock-in, and inequities of technology access, especially for broadband internet, among students. The authors noted that a lack of adoption models appeared to diminish the administrators' confidence in open-source software, and smaller institutions were more likely to favor piloting open-source tools. They also found that institutions with stable enrollments were more likely to consider the effects of low-cost technology solutions on student perceptions, relative to institutions with student retention challenges.

This research emphasizes adoption as a key administrative return-on-investment (ROI) metric. It also notes that differences between stakeholder priorities and between technology needs of course cohorts are often revealed and tested at scale. Our contribution to this discussion is a vendor-agnostic adoption measurement and claim-testing strategy applicable to any number of courses, staff-favored apps and support strategies. 

\begin{table}[t]
\caption{A taxonomy of frequent use-contexts for a learning management system, drawn from instructional designers' experience of course redesign initiatives at Virginia Tech.}
\centering
\label{tab_taxonomy}
\begin{tabular}{p{4cm} p{9cm}}
\toprule
\bfseries LMS Resource & \bfseries Use Context\\
\midrule
Announcements (\textbf{An}) & \textbf{0}: None; \textbf{1}: Placeholder announcements; \textbf{2}: At least one per week or course instrument\\

Syllabus (\textbf{S}) & \textbf{0}: None; \textbf{1}: Syllabus under \textit{Files}; \textbf{2}: File previewed/embedded under \textit{Syllabus}\\

Discussions (\textbf{D}) & \textbf{0}: Discussions disabled; \textbf{1}: No discussion activity; \textbf{2}: One or more live discussion threads (at least one post per week or course instrument)\\

Assignment Delivery ($\textbf{A}_{d}$) & \textbf{0}: No assignments on LMS or placeholders; \textbf{1}: Link to DOC, ZIP or 3rd-party app; \\

 & \textbf{2}: Assignments fully hosted on LMS\\

Quiz Delivery ($\textbf{Q}_{d}$) & \textbf{0}: No assignments on LMS or placeholders; \textbf{1}: Link to DOC, ZIP or 3rd-party app; \\

 & \textbf{2}: Quizzes fully hosted on LMS\\
 
Assignment Submission ($\textbf{A}_{s}$) & \textbf{0}: No file upload, likely paper or 3rd-party app; \textbf{1}: LMS file upload; \textbf{2}: LMS text entry\\

Quiz Submission ($\textbf{Q}_{s}$) & \textbf{0}: No online submission, likely paper or 3rd-party app; \textbf{1}: Submission within LMS\\

Gradebook (\textbf{G}) & \textbf{0}: No grading activity in LMS; \textbf{1}: Comprehensive grading for all assessments\\

Files (\textbf{F}) & \textbf{0}: No files; \textbf{1}: Course resources under \textit{Files}\\

\bottomrule 

\end{tabular}
\end{table}

\subsection{Educational data mining and learning analytics}
A discussion of the key drivers of learning analytics research in \citet{ferguson2012learning} and \citet{dawson2010seeing} notes how native LMS data analysis, visualization and recommendation capabilities are presently non-existent or quite limited, even with standard tracking software features. A lot of student activity is external to the LMS, the data volume is huge and ever-expanding, and there is little standardization of the data aggregation and reporting methods, viz-a-viz critical use-cases for all stakeholders involved (faculty, students, instructional designers, LMS administrators, department leadership). These problems persist even as in the past two decades, inroads in educational data mining \citep{romero2008data, romero2010survey, sdmtutorialhuzefa} have helped advance the state of the art in predictive modeling of student engagement, learning, and achievement \citep{engagementhenrie2015, black2008data, cocea2006can, cocea2007cross}. Simultaneously, LMS log data analyses have been used extensively to model student and faculty use-contexts \citep{casany2012analyzing, mazza2004gismo}, and to improve LMS features \citep{fenu2017learning}, often for specific disciplines and pedagogies \citep{hassan2020iticse}. 
Improving existing pedagogies \cite{williams2023data}, assessing learning outcomes and risk-of-failure for students \citep{he2015aaai, elbadrawy2015collaborative}, and recommending interventions are all important use-cases that call for a convergence of data sources, a synthesis of approaches, and consensus among stakeholders. One of the early instances of this approach is Course Signals at Purdue \citep{arnold2012course}. Course Signals uses students' course outcomes, frequency of interaction with the LMS (Blackboard Vista), prior academic history and demographic information to ascertain a failure-risk measurement. In \cite{wolff2013improving}, a short-term warning system for ailing students models the early-term drop in clickthrough rates for modules of an online course. \cite{macfadyen2010mining} describe a similar early-warning system which identifies isolated students using an analysis of ego networks and micro-communities of high-ability students on an online course forum.

The breadth of qualitative correlates of LMS adoption reviewed in prior research highlights how complex - and potentially useful - it is to assign context to LMS data. A variety of stakeholders (figure \ref{fig_sys_overview}) bring competing standards to evaluate the quality of the content delivered via LMS course sites. This suggests the need for a thorough, quantitative, and scalable means of evaluating LMS use by resource and context (table \ref{tab_taxonomy}). To the best of the authors' knowledge, our work contributes a first formal, fine-grained, and vendor-agnostic method of measuring user engagement and discovering micro-cohorts of courses aboard an LMS. 

\section{Depth of Use: Requirements, Estimation, Study Objectives} 
In this section, we identify the design constraints of a multi-service user-engagement metric for a learning management system (LMS). We then define a service-level LMS depth-of-use ($DOU$) and describe how multiple service-level $DOU$s can be aggregated into a single course-level DOU. We also describe four research questions (and ten corresponding hypotheses) which test how strongly DOU for a course is correlated with its modality, participation, logistics and outcomes.

\begin{figure}[!t]

\begin{subfigure}{1.0\textwidth}
\footnotesize
\centering
\caption{}
\begin{tabular}{|c|c|c|c|}
\hline
\bfseries  & \bfseries  & \bfseries Weekly Page Requests & \bfseries Weekly Page Requests\\
\bfseries Course Code & \bfseries Course Name & \bfseries (Non-Participants) & \bfseries (Participants)\\
\hline
CEE 1234 & Theory of Structures & 266.92 & 67.1\\
PSYC 1234 & Psychology of Learning & 51.6 & 72.8\\
BIOL 1234 & General Microbiology & 176 & 17.48\\
PSYC 1244 & Nervous System and Behavior & 54.7 & 30.52\\
\hline 
\end{tabular}
\end{subfigure}

\begin{subfigure}{1.0\textwidth}
\centering
\caption{}
\includegraphics[width=0.4\textwidth]{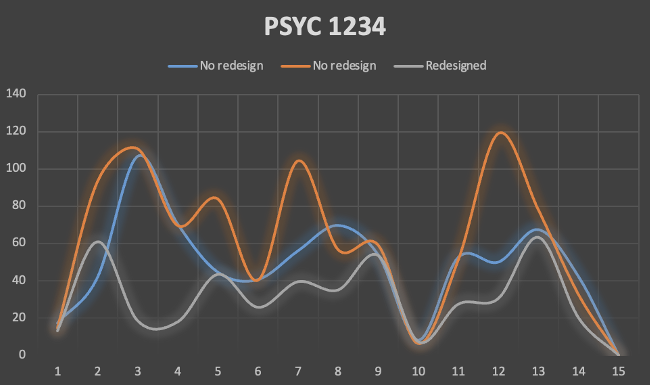}
    \qquad
    \footnotesize
    \begin{tabular}[b]{|c|c|}\hline
    \multicolumn{2}{|c|}{\textbf{Redesigned Course: Page Requests}}\\ \hline
      Type, Action & \% \\ \hline
      course, activity stream summary & 26 \\
      course, show & 23.3 \\
      folders, resolve path & 8.5 \\
      file previews, show & 6.6 \\
      folders, show & 4.7 \\
      files, index & 3.7 \\ 
      gradebook, summary & 3 \\
      submissions, update & 3 \\
      discussions api, view & 3 \\
      discussion topics, show & 3 \\
      \hline
    \end{tabular}
    \captionlistentry[table]{A table beside a figure}
\end{subfigure}

\begin{subfigure}{1.0\textwidth}
\footnotesize
\centering
\caption{}
    \begin{tabular}[b]{|cc|ccc|}\hline
    \multicolumn{2}{|c|}{\textbf{Fall 1234 Department Overview}} & \multicolumn{3}{c|}{\textbf{Page Requests Motifs by Department}}\\ \hline
      Department & \# Courses & Department & Page Requests & \%\\ \hline
      ENGL & 224 & ISE & course (activity summary, show), submissions api & 49\\
      MATH & 169 & ISE & course (activity summary, show), folders (resolve path) & 42\\
      ECE & 130 & ISE & files (api), courses (show, activity summary) & 37\\
      BIOL & 110 & ISE & courses (activity summary, show), folders (show) & 32\\
      CEE & 99 & ISE & courses (show, activity summary), assignments (syllabus) & 31\\
      \hline
    \end{tabular}
    \captionlistentry[table]{A table beside a figure}
\end{subfigure}

\begin{subfigure}{1.0\textwidth}
\footnotesize
\centering
\caption{}
    \begin{tabular}[b]{|c|c|c|c|}\hline
    \multicolumn{4}{|c|}{\textbf{Canvas Use: Spring 1234 Course Cohort}}\\ \hline
      Category & Criteria & \# & \% \\ \hline
      Overall & any announcements, discussions, OR assignments & 3578 & 94.6 \\
      High & announcements, discussions, AND fully-hosted assignments & 319 & 8.4\\
      High & (announcements OR discussions) AND fully-hosted assignments) & 1319 & 34.8\\
      High & announcements OR discussions, OR fully-hosted assignments & 2797 & 73.9\\
    Medium & (announcements OR discussions) AND assignment hosting (DOC/ZIP) & 726 & 19.2\\
    Medium & assignment hosting (DOC/ZIP) & 929 & 24.5\\
    Low & no announcements, no discussions, AND no assignments & 203 & 5.3\\

      \hline
    \end{tabular}
    \captionlistentry[table]{A table beside a figure}
\end{subfigure}

\caption{Brainstorming session aids for Canvas depth-of-use development: three early, data-driven attempts to capture Canvas use from page request logs ((a)-(c)), and a preliminary model-driven summary of Canvas usage tiers (d).}
\label{fig_dou_dev}
\end{figure}


\subsection{Design constraints and requirements}
The study began with a series of informal brainstorming sessions with the program administrators: three instructional designers, a director of learning experience design, and a director of IT software development. The team sought to understand the challenges of evaluating the effectiveness of a course development program offered by the division of IT at Virginia Tech. This initiative was structured as a semester-length course with faculty-designer interest groups meeting weekly to design a new Canvas course, and building competency in topics of active learning, self-paced modules, flipped classrooms, lecture capture, accessibility, and copyright and fair use. Course faculty enrolled in the initiative worked on weekly assignments targeting syllabus review, assessments, online pedagogy, student and classroom management, and alignment with learning outcomes. Our interviewees expressed significant interest in a course-level metric of user engagement with Canvas as a first step in evaluating faculty's reuse of best practices from this program, a ``performance scorecard'' for the program (figure \ref{fig_dou_dev}). Weekly and monthly page request counts from Canvas logs (normalized by course enrollment) were the initial metric of choice, followed by page request compositions and semester-length timelines. The author recorded notes on their laptop during the sessions. The administrators' discussions centered on whether the metric adequately highlighted the differences between cohort participants and non-participants, and could facilitate conversations between designers and Virginia Tech's faculty clients - with a broad range of quantitative competencies - on how best to achieve this improvement in student engagement. Figure \ref{fig_dou_dev}a through \ref{fig_dou_dev}c illustrate the use of weekly course page requests to highlight the distinctions between program participants and non-participants, and identify popular Canvas services. 

Two key requirements for this metric emerged from our brainstorming sessions: the need for an \textbf{ordinal engagement metric} instead of a continuous-valued one, and the flexibility to incorporate \textbf{the use of new tools and services}. Instructional designers noted the challenges in using a ``percentage engagement" metric to help their faculty clients infer a meaningful target LMS engagement: defining a meaningful baseline, identifying the native LMS and third-party apps most relevant to the course, and enabling quick inference of best practices for faculty with a variety of quantitative skills. They also noted a variety of legacy tools in use at different Virginia Tech colleges and departments, and expressed the need to account for potential disparities in uptake rates of new, LMS-compatible tools. LMS "usage tiers" (figure \ref{fig_dou_dev}d) thus provided an effective design alternative.

These considerations critically inform our ordinal, modular formulation of an LMS ``Depth of Use'' metric. In our sessions, we solicited a simple tabulation of the designers' mental models of Canvas use, which we evaluated for consensus, consistency, and accuracy. Table \ref{tab_taxonomy} outlines the resulting taxonomy of ``low'', ``medium'' and ``high'' use of seven Canvas services (announcements, syllabus, discussions, assignments, quizzes, gradebook, and files). This taxonomy forms the basis of course-level DOU estimation in our study.

\begin{figure*}
\centering
\includegraphics[width=0.9\textwidth]{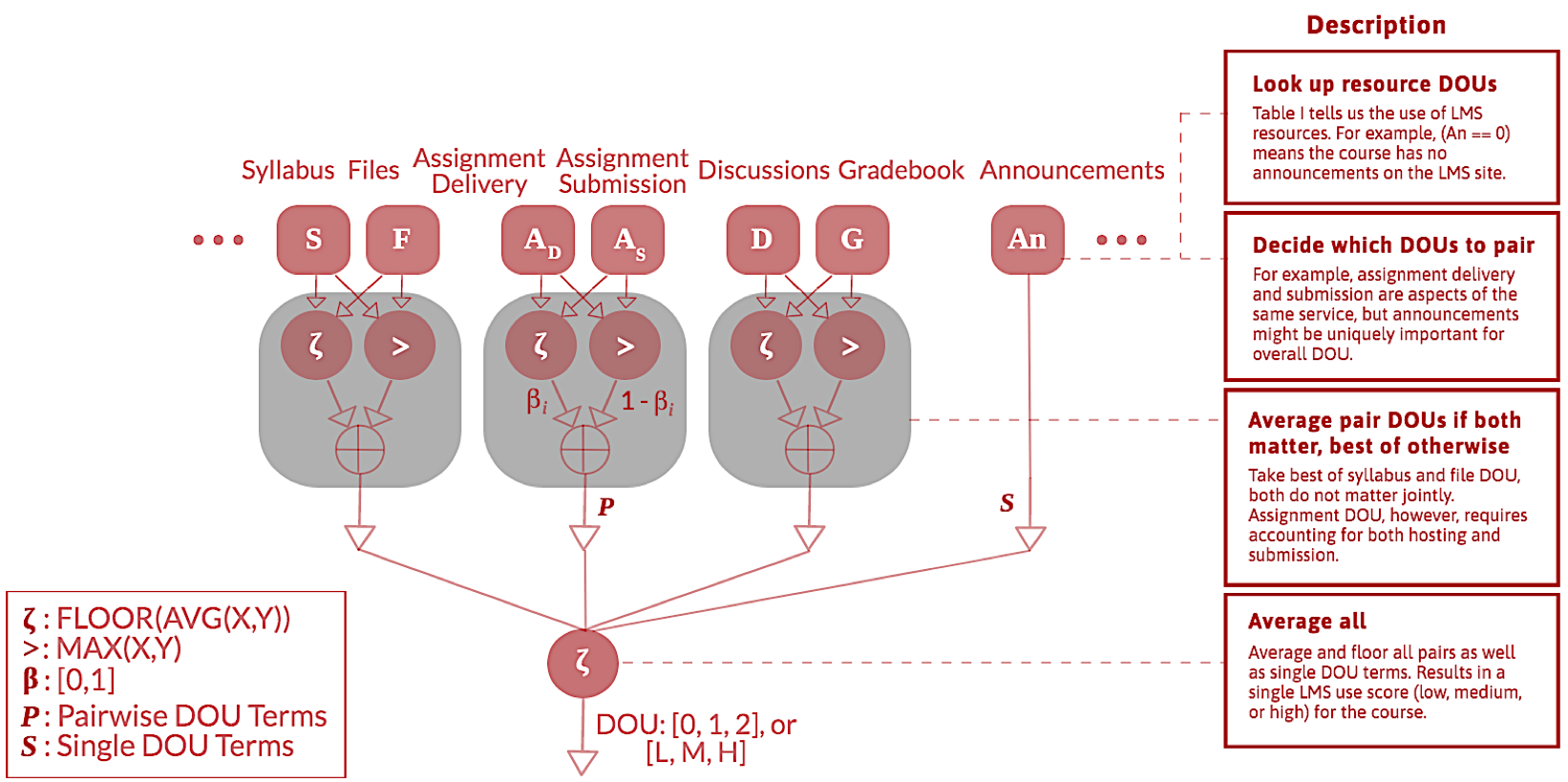}
\caption{A schematic (left), and descriptions of all steps (right) involved in the calculation of course-level DOU. S, F, D, etc. refer to LMS resource labels in the taxonomy in table \ref{tab_taxonomy}. These resource-level DOUs are aggregated into the overall DOU ranking (low, medium, or high) for the course.}
\label{fig_dou_calc}
\end{figure*}

\subsection{Notation and definitions}
We define depth-of-use for an LMS service or resource $R_{i}$ as a simple logic rule $DOU_{i}$ of the form $(R == k_{i})$ where $k_{i}$ is a whole number. For instance, per table \ref{tab_taxonomy}, $(\textbf{An}==1)$ for a given course implies \textit{some} use of announcements (placeholders or class schedules, no instructor or TA activity). A total of $N$ resource $DOU$s are accounted towards each course. As visualized in figure \ref{fig_dou_calc}, the overall DOU for the course, DOU$_{c}$ is \textit{aggregated} from the resource \textit{DOU}s as follows. $P$-terms refer to pairs of LMS resource \textit{DOU}s, and $S$-terms refer to single resource \textit{DOU}s.

\begin{equation}
\label{course_dou}
\text{DOU}_{C} \triangleq \zeta (P_{1}, P_{2}, .. P_{M'}, S_{1}, S_{2}, .., S_{N'})
\end{equation}

where 

\begin{equation}
\label{main_dou_equation}
P_{i}^{(A,B)} = (1-\beta_{i}) \bigg(MAX \Big(DOU_{A}, DOU_{B} \Big) \bigg) +
\beta_{i} \bigg(\zeta \Big(DOU_{A},DOU_{B} \Big) \bigg)
\end{equation}

\begin{figure*}
\centering
\includegraphics[width=0.6\textwidth]{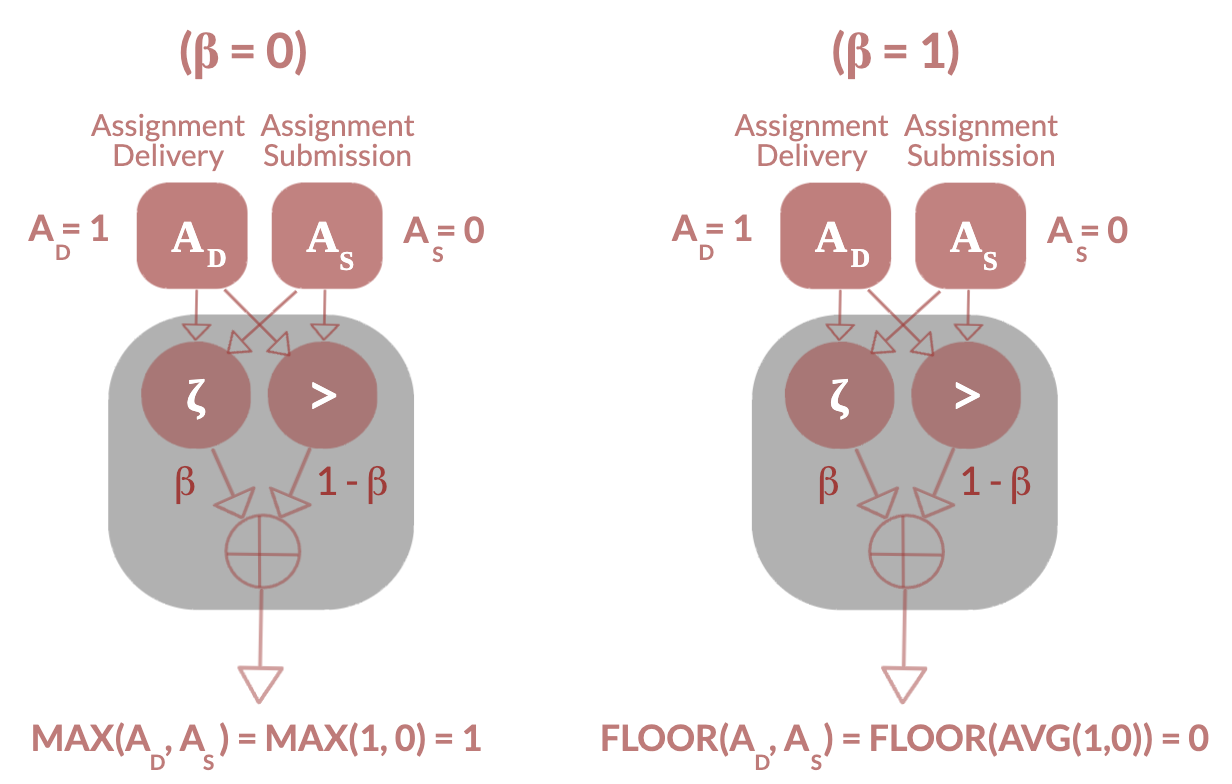}
\caption{A simple illustration of how two LMS resource DOUs are paired using equation \ref{main_dou_equation}. In this example, assignment delivery \textit{DOU} is 1 (link to DOC, ZIP or app), and assignment submission \textit{DOU} is 0 (no file upload, likely paper or app). (Left) Setting $\beta$ to 0 computes the MAX of the two resource \textit{DOU}s, which is useful when only one is needed for overall DOU. (Right) Setting $\beta$ to 1 computes the floored average, which is useful when both need to contribute to DOU equally.}
\label{fig_dou_pair}
\end{figure*}

Equation \ref{main_dou_equation} describes how two resource \textit{DOU}s \textit{A} and \textit{B} are paired in $P_{i}$. We choose to apply MAX() or $\zeta()$ by setting $\beta_{i}$ to 0 or 1, respectively. $\zeta()$ is the logic equivalent of a real-valued floored-average $\floor{AVG(X,Y)}$ function. Figure \ref{fig_dou_pair} illustrates this step for assignment delivery $(\textbf{A}_{d}==1)$ and assignment submission $(\textbf{A}_{s}==0)$. $MAX()$ assigns the output to the larger of the two input contributions, hence $P_{i}^{(\textbf{A}_{d}, \textbf{A}_{s})}=1$. On the other hand, $\zeta()$ gravitates to the lower of the two, hence $P_{i}^{(\textbf{A}_{d}, \textbf{A}_{s})}=0$. Picking $\beta_{i}=0$ implies that the instructional staff intends to consider the MAX(), or the \textit{best} of assignment delivery and submission $DOU$s towards the overall LMS DOU. On the other hand, $\beta_{i}=1$ rewards contributions from both DOUs equally, so \textit{both} assignment delivery and submission need to utilize Canvas thoroughly for a high overall DOU rank. Finally, in equation \ref{course_dou}, we average all of the pairwise ($P$), and single ($S$) terms using $\zeta()$ to compute a final score (low, medium or high) of overall LMS use. The findings in Section 4 are based on $\beta_{i}=1$ for pairwise DOU terms ($\textbf{A}_{d}$, $\textbf{A}_{s}$) and ($\textbf{Q}_{d}$, $\textbf{Q}_{s}$), and $\beta_{i}=0$ for pairwise DOU terms (\textbf{S}, \textbf{F}), and (\textbf{D}, \textbf{G}). Note that DOU allows flexibility in both \textit{pairings} and \textit{weights}, to encourage research on the usability and perceived efficacy of custom DOUs for a variety of tools and learning environments. In addition, table \ref{tab_taxonomy} is vendor-agnostic, in that it can measure the use of multiple LMS ecosystems, and taxonomies for LMS resources can be added or subtracted on a need-basis. We identify these design vectors in our discussion of future work (see Section 6). 

Three important practical considerations emerge in the design of DOU. First, courses frequently contain multiple delivery and submission types for assignments and quizzes. We thus require the overall DOU criteria in the taxonomy to hold true for a simple majority (at least 50\%) of assignments or quizzes in the course. For instance, at least 50\% of the course assignments should be fully hosted on Canvas for the assignment delivery \textit{DOU} to assume a value of 2. Second, we define high discussions \textit{DOU} ($\textbf{A}_{d}$) with the presence of one or more live Canvas discussion threads (at least one post per week or course instrument), same as the announcements \textit{DOU} ($\textbf{An}$). These heuristics aid the overall parsimony and interpretive power of the DOU taxonomy. Third, we do not incorporate issues of information quality and use-quality for specific LMS tools, such as relevance and degree of reflection in discussion posts, or the ease-of-use and diversity of assignment submission modalities (notebooks, error logs, images, hyperlinks). These aspects are important to evaluate in faculty's software use. However, they pose significant challenges related to data sparsity (wide variation in data availability across departments), need for domain knowledge (learning theories, objects, and environments in use), and lack of feature parity across LTI apps, making the design of a unified taxonomy extremely challenging. DOU is envisioned foremost as a platform-level metric of LMS use across departments, so we identify issues of information-quality and use-quality as outside the scope of the current taxonomy, and reserve them for future work.

\subsection{Research questions and hypotheses}
In this section, we describe our study research questions (\textbf{RQ1}, \textbf{RQ2}), and ten corresponding hypotheses (\textbf{H1-H8} and \textbf{H9-H10}, respectively). These hypotheses test how significant the connection is between DOU and course attributes like modality (\textbf{H1} - \textbf{H4}), participation (\textbf{H5}, \textbf{H6}), outcomes (\textbf{H7}, \textbf{H8}), and logistics (\textbf{H9}, \textbf{H10}). Note that these hypotheses can inform both research questions, and we discuss these connections in Section 5. 


\paragraph{\textbf{RQ1} (\textbf{CI}, adoption contexts)} What aspects of the course design, content and delivery drive the faculty and students to utilize or ignore individual LMS tools?

\begin{itemize}
    \item \textbf{H1:} Undergraduate courses have significantly higher DOUs relative to graduate DOUs.
    \item \textbf{H2:} STEM courses have significantly higher DOUs relative to non-STEM courses.
    \item \textbf{H3:} Online-only courses have significantly higher DOUs relative to face-to-face courses.
    \item \textbf{H4:} Third-party app use significantly affects DOU.
    \item \textbf{H5:} Course DOU is significantly linked to the number of students enrolled full-time in the course.
    \item \textbf{H6:} Course DOU is significantly linked to pageviews for the LMS course website.
    \item \textbf{H7:} Course DOU is significantly linked to the average GPA awarded in that course.
    \item \textbf{H8:} Course DOU is significantly linked to the DFW rate of that course.
\end{itemize}

\paragraph{\textbf{RQ2} (\textbf{ORG}, institutional support)} How do we produce quantitative insights to inform institutional support (teacher support, professional development, pandemic response)?

\begin{itemize}
    \item \textbf{H9:} Course DOU is significantly linked to the number of teaching staff members for the course.
    \item \textbf{H10:} Course DOU is significantly linked to the instructor's prior enrollment in on-demand coursework and training.
\end{itemize}

\begin{table}[t]
\caption{Key counts and DOU breakdown (\% \textbf{Lo}, \textbf{Med}, \textbf{Hi}) for course cohorts in the spring 2023 dataset.}
\centering
\label{tab3}
\begin{tabular}{ccc}
\toprule
\bfseries Attribute & \bfseries \# & \bfseries \%\\
\midrule
Overall & 4866 & 21, 61, 18\\
Undergraduate & 3825 & 18, 62, 20\\
STEM & 2982 & 22, 62, 16\\
Online & 1519 & 41, 43, 16\\
Third-Party App Use & 228 & 2, 74, 24\\
Digital Skills Training & 2781 & 21, 60, 18\\
Viewership ($\mu, \sigma$) & 773, 657 & -\\
Enrollment ($\mu, \sigma$) & 51, 102 & -\\
\#TAs ($\mu, \sigma$) & 0.9, 2.9 & -\\
\bottomrule 

\end{tabular}
\end{table}

\begin{table}[t]
\caption{Spring 2023: High, medium and low DOU group composition (\%) by course and instructor attributes.}
\centering
\label{tab_dou_comp}
\begin{tabular}{cccccc}
\toprule
\bfseries DOU & \bfseries Undergrad & \bfseries STEM & \bfseries Online & \bfseries App use & \bfseries Skills\\
\midrule
Low & 66 & 64 & 62 & 0.4 & 59\\
Medium & 80 & 63 & 22 & 5 & 57\\
High & 84 & 53 & 26 & 6 & 56\\
\bottomrule 

\end{tabular}
\end{table}

\section{Evaluation}
\subsection{Datasets}
The primary dataset for this study is metadata collected for 35037 courses offered between spring 2021 and spring 2023 from Canvas, the enterprise LMS in operation at Virginia Tech. Table \ref{tab3} lists key aspects of the 4866 courses analyzed during the spring of 2023. For instance, 3825 (78.6\%) of these courses are intended for undergraduate audiences, 2982 (61.2\%) courses deal with STEM content, and 1519 (31.2\%) are virtual offerings. These majorities are also retained in each of the three DOU groups as per table \ref{tab_dou_comp}, with important differences. Section 6 discusses these patterns in detail. We used a combination of manual and automated strategies (web scraping, entity resolution, and topic modeling) to create LMS utilization metadata for each course. Key textual sources include, and are not limited to, the Virginia Tech course catalog and historical timetable, Canvas page request logs, course descriptions on the Virginia Tech (TLOS Professional Development Network) website \footnote{https://profdev.tlos.vt.edu/}, as well as syllabus files and assessment page content from Canvas course sites. STEM tagging of courses in the dataset is in accordance with the DHS classification of STEM fields \citep{stemlist2012}. We also conducted two rounds of semi-structured collaborative sensemaking sessions with Virginia Tech instructional designers, software developers, and faculty. The first round (N=4) focused on discovering design requirements for DOU and synthesis of the designers' mental models towards the DOU source taxonomy (table \ref{tab_taxonomy}). The second round (N=7) focused on expert reviews of DOU to discover actionable low-adoption cohorts (Section 5.2).

\subsection{Methods}
To answer our research questions (section 3.3), we begin by testing our fundamental hypotheses (\textbf{H1} - \textbf{H10}). DOU is ordinal and not normally distributed, so we use non-parametric Kruskal-Wallis H-test \citep{kruskal}, in addition to an independent two-sample t-test, for hypotheses with discrete-valued meta-variables (Table \ref{tab_hyp}). We evaluate group differences in viewership and enrollment for each of low, medium and high DOUs using one-way ANOVA (F-test, Table \ref{tab_hyp}). To expand our analysis, we then test each of the hypotheses (ANOVA: table \ref{tab_hyp_indiv}) against all constituent dimensions of DOU. In Section 5, we combine these hypothesis tests with frequency and cohort analyses, in accordance with the needs of the DOU use-case (adoption, institutional support).
\begin{table}[t]
\caption{Spring 2023: Hypothesis-testing the relationship between LMS DOU and key course attributes.}
\centering
\label{tab_hyp}
\begin{tabular}{cccc}
\toprule
\bfseries Hypothesis & \bfseries $t$ & \bfseries $F$ & \bfseries $H$\\
\midrule
\textbf{H1}: Undergraduate & 9.8** & 96** & 95**\\
\textbf{H2}: STEM & -5* & 25.1* & 24.7*\\
\textbf{H3}: Online & 7.1** & 50.2** & 51**\\
\textbf{H4}: App use & 5.9* & 34.6* & 34.8*\\
\textbf{H5}: Enrollment & - & 85.7** & 1e3**\\
\textbf{H6}: Viewership & - & 9.7e2** & 2e3**\\
\textbf{H7}: GPA & - & 28.9** & 75.7**\\
\textbf{H8}: DFW & - & 15.6* & 77**\\
\textbf{H9}: \#TA & - & 67.6** & 4.7e2**\\
\textbf{H10}: Skills & -1.36 & 1.84 & 1.85\\
\bottomrule
\multicolumn{3}{l}{*stat. signif., $\alpha=0.05, p<=\alpha \land $ $p>1e{-10}$, **$p<1e{-10}$}\\
\multicolumn{3}{l}{*$F>F_{crit}$, $F_{ov} = 3.5^{*}$}\\

\end{tabular}
\end{table}

\subsection{Results}
\subsubsection{Modality (\textbf{H1-H4})} As per Table \ref{tab_hyp}, undergraduate courses have higher average DOUs relative to graduate courses (t-statistic is positive), consistent with their higher average enrollment (60 as opposed to 20 for graduate courses). As per table \ref{tab_hyp_indiv}, undergraduate courses have higher relative DOUs for announcements ($F=164.8$**), grading ($F=169$**) and quiz submission ($F=196$**), among others. Non-STEM courses feature higher use of the LMS for assignment delivery ($F=27.3$*) and submission ($F=7.3$*), among others. Traditional in-class instruction loses out to online-only courses in overall DOUs. Online instruction is linked to in-depth use of online syllabi ($F=49.8$**), as well as assignment delivery ($F=10.8$*) and submission ($F=18.5$*). Reliance on third-party apps coincides with the use of announcements ($F=71.4$**), gradebook ($F=25.7$*) and discussion forums ($F=11.5$*), among others.

\subsubsection{Participation (\textbf{H5-H6})} Higher \textit{DOU} courses feature larger overall enrollment ($F=85.7$**) and viewership ($F=9.7e2$**), as per table \ref{tab_hyp}. Both of these are strong correlates of LMS utilization overall, and across a number of LMS resources considered individually (table \ref{tab_hyp_indiv}). High enrollment is linked to high use of detailed online announcements ($F=114$**), assignment delivery ($F=34.8$**) and discussion forums ($F=24.1$**), among others. High site viewership is similarly linked to the use of syllabi ($F=215$**), assignment delivery ($F=738$**) and gradebook ($F=1e3$**), etc.

\subsubsection{Outcomes (\textbf{H7-H8})} The average course GPA is significantly linked to overall DOU as per table \ref{tab_hyp} ($F=28.9$**), and the use of announcements ($F=73.2$**), syllabi ($F=19.7$*) and discussion forums ($F=37.8$**), among others. DFW rate is also a correlate of DOU ($F=15.6$*). Smaller DFW rates coincide with higher online quiz submission ($F=23.2$**), announcements ($F=37.5$**) and gradebook ($F=27$*), among others. It is important to note the complexity of assessing LMS engagement in its impact on course outcomes without multivariate analyses and thorough accounting of confounding variables (see Section 6 for associated directions for future work).   

\subsubsection{Logistics (\textbf{H9-H10})} As per table \ref{tab_hyp_indiv}, the number of teaching assistants is significantly linked to higher DOUs for announcements ($F=83.4$**), discussion forums ($F=7$*) and gradebook ($F=86.8$**). Participation in an online digital skills training program run by Virginia Tech is not strongly linked to overall LMS use. It is nonetheless linked to higher resource DOUs for discussion groups, syllabi, and files.

\begin{table}[t]
\centering
\caption{Hypothesis tests: $t$ and $F$ statistics for the relationship between resource DOU and course attributes.}

\begin{tabular}{ccccc}
\toprule
\bfseries Hypothesis & \bfseries An & \bfseries S & \bfseries F & \bfseries $A_{d}$\\
\midrule
\textbf{H1}: Undergraduate & 12.8**, 164.8** & 2.5*, 6.2* & 8.3**, 69.2** & 11.9**, 141.3**\\
\textbf{H2}: STEM & 1.99*, 3.97* & -6.7**, 45.6** & 0.92, 0.84 & -5.2*, 27.3*\\
\textbf{H3}: Online & 0.7, 0.49 & 7**, 49.8** & 1, 1 & 3.3*, 10.8*\\
\textbf{H4}: App use & 8.4**, 71.4** & 3.2*, 10.5* & 7.3**, 53.1** & 7.3**, 53.1**\\
\textbf{H5}: Enrollment & - , 114** & - , 15.9* & -, 141.6** & -, 34.8**\\
\textbf{H6}: Viewership & - , 727.4** & - , 215* & -, 2e3** & -, 738**\\
\textbf{H7}: GPA & - , 73.2** & - , 19.7* & -, 145.7** & -, 16.6*\\
\textbf{H8}: DFW & - , 37.5** & - , 16.5* & -, 57.9** & -, 6.5*\\
\textbf{H9}: \#TAs & - , 83.4** & - , 15* & -, 105** & -, 30.3**\\
\textbf{H10}: Skills & -0.6, 0.4 & -0.02, 0 & -1.97*, 3.8* & -1.3, 1.8\\

\bottomrule
\multicolumn{5}{l}{*stat. significant, $\alpha=0.05, p<=\alpha \land $ $p>1e{-10}$, **$p<1e{-10}$, $F>F_{crit}$}\\
\end{tabular}

\vspace{10pt}
\begin{tabular}{cccccc}
\toprule
\bfseries Hyp. & \bfseries $A_{s}$ & \bfseries $Q_{d}$ & \bfseries $Q_{s}$ & \bfseries G & \bfseries D\\
\midrule
\textbf{H1} & 13.1**, 171.5**  & 12.2**, 150** & 14**, 196** & 13**, 169** & -0.75, 0.56\\
\textbf{H2} & -2.7*, 7.3* & 5.9*, 35* & 5.8*, 34.5* & 1.2, 1.4 & -18**, 328**\\
\textbf{H3} & 4.3*, 18.5* & 3.5*, 12.7* & 1, 1 & 1.5, 2.4 & 15**, 221.8**\\
\textbf{H4} & 8.9**, 80.6** & 9**, 81.2** & 9.6**, 92** & 5*, 25.7* & 3.4*, 11.5*\\
\textbf{H5} & - , 76**  & -, 135.5** & -, 238** & -, 98.3** & -, 24.1**\\
\textbf{H6} & - , 1e3** & -, 706** & -, 1e3** & -, 1e3** & -, 230**\\
\textbf{H7} & - , 25**  & -, 55.8** & -, 146** & -, 125** & -, 37.8**\\
\textbf{H8} & - , 4.9*  & -, 9.3* & -, 23.2** & -, 27* & -, 23.4**\\
\textbf{H9} & - , 64** & -, 95.4** & -, 173** & -, 86.8** & -, 7*\\
\textbf{H10} & -1.7, 2.9 & -1.2, 1.6 & -1, 1.1 & -2.9*, 8.6* & -2.1*, 4.5*\\

\bottomrule
\multicolumn{5}{l}{*stat. significant, $\alpha=0.05, p<=\alpha \land $ $p>1e{-10}$, **$p<1e{-10}$, $F>F_{crit}$}\\
\end{tabular}
\label{tab_hyp_indiv}
\end{table}

\section{Applications and Discussion}
Having surveyed the relationship between DOU and key course characteristics (modality, participation, logistics and outcomes), we discuss three applications of our analyses. We begin by describing how faculty members can use DOU to understand the utility of LMS services relative to legacy apps and the opportunity-cost of a future transition. We then describe how DOU can evaluate the efficacy of professional development programs and resource allocation at the department level. Finally, we describe how LMS administrators can use DOU to look for actionable low-adoption micro-cohorts of courses.

\subsection{(RQ1) What drives faculty to adopt or ignore LMS tools?}
The literature reviewed in section 2 and the outcomes of our analyses in section 4 reflect the multifaceted roots of LMS adoption. These factors potentially include perceived quality (system, use, information), administrative decisions (rollout strategies, teaching assistance, technical support, LMS evangelism), departmental precedents, and end-user attributes (technology self-efficacy, leadership, work objectives and career priorities). Ascertaining the role of these factors remains an open question, but our numerical evaluation points us to faculty needs of scale, interoperability, and ubiquitous access. We review these needs as follows:

\begin{figure}
\centering
\includegraphics[width=0.5\textwidth]{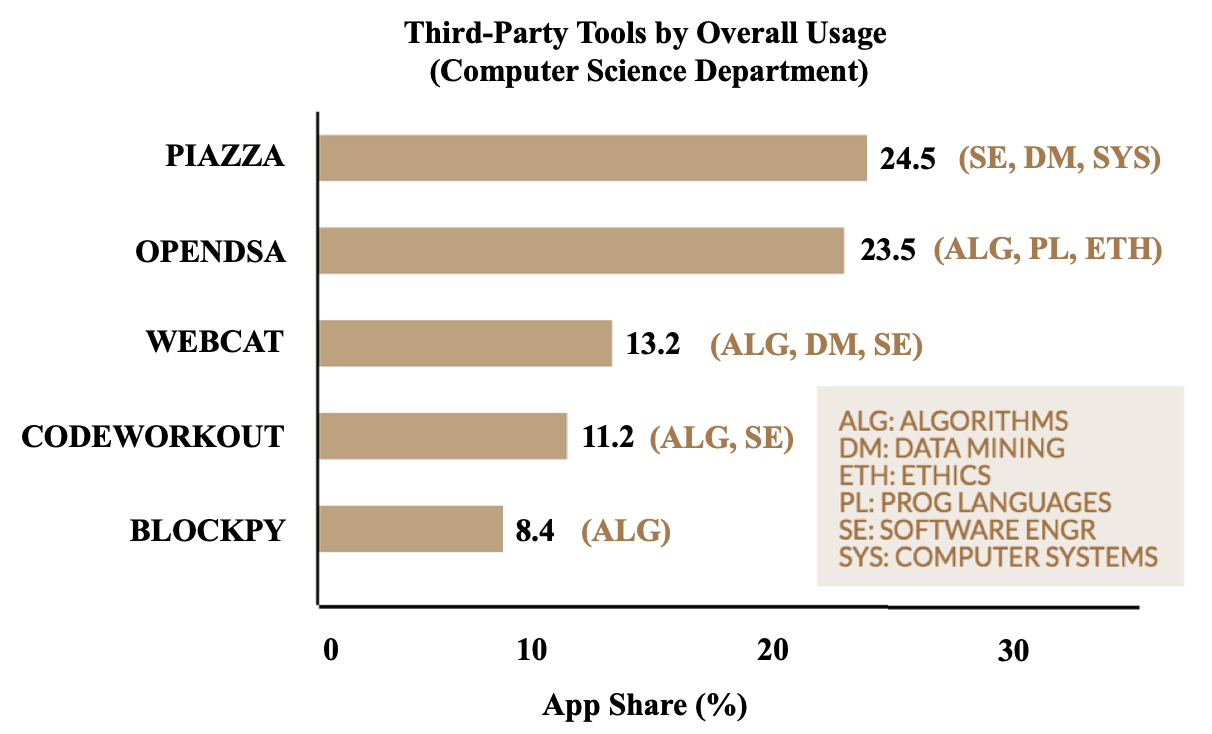}
    \qquad
    \small
    \begin{tabular}[b]{|c|c|}\hline
    \multicolumn{1}{|c|}{\textbf{Third-Party Tools by Overall Usage}}\\ 
    \multicolumn{1}{|c|}{\textbf{(Across Departments)}}\\     
    \hline
      EDU-APPS\\
      WEBASSIGN\\
      WILEY\\
      APPORTO\\
      BADGER\\
      PIAZZA\\ 
      SCORM\\
      SMARTSPARROW\\
      ELIREVIEW\\
      PEARSONCMG\\
      \hline
    \end{tabular}
    \captionlistentry[table]{A table beside a figure}
    \caption{Third-party app use at Virginia Tech by frequency of use, overall (right) and in the Computer Science department (left). Discussion forums and course content management (Piazza), and programming instruction (OpenDSA) combined account for nearly half of all third-party app use for the latter. Significant, often LMS-unaccounted use of third-party apps makes it challenging for department and IT leadership to evaluate and scale pedagogies.}
\label{fig_app_share}
\end{figure}



\subsubsection{Managing Scale} As per hypothesis \textbf{H1} in table \ref{tab_hyp_indiv}a, larger class size coincides with higher or `deeper' use of announcements, most likely because mailing lists become increasingly inefficient and harder to organize and search at scale. Larger audience sizes also coincide with more frequent LMS use for assignment submission and delivery. One key reason is that this allows for a larger range of content to be submitted and greater flexibility in scheduling and organizing take-home exams and offline evaluations. According to hypothesis \textbf{H4} in table \ref{tab_hyp_indiv}, the use of third-party apps coincides somewhat weakly with the use of discussion forums. Services like Piazza are particularly favored by faculty because of their advanced forum management, content processing and tagging features, compared to the newer Discussions app aboard Canvas. The largest effect sizes for \textbf{H5} (table \ref{tab_hyp_indiv}) correspond to the use of quiz submission, quiz delivery, announcements, and gradebook. Undergraduate, non-STEM courses are likely to utilize these LMS tools. These courses are typically major-unrestricted, and enroll hundreds of students across multiple sections in a given academic term. Hypothesis \textbf{H9} (tables \ref{tab_hyp_indiv}) suggests this also coincides with higher numbers of teaching assistants. Early adopters in the instructional staff of these courses especially gravitate towards basic \textit{housekeeping} use-cases for LMS tools, such as communicating class times, office hours, course milestones, and grades, whilst retaining their use of third-party apps (\textbf{H4}). Faculty's ability to delegate administrative and technology discovery tasks can thus critically help them balance their research and teaching duties and potentially migrate to new tools as class sizes increase. Third-party apps with free tiers, local authorship, and open-source communities remain consistently popular because of their low overhead of initial setup. But, without adequate access to teaching assistance, scaling the use of these apps to high-enrollment classes, managing student feedback, and providing timely technical support are likely to remain challenging. 

\subsubsection{Ensuring Interoperability and Ubiquitous Access} Intuitive, safe, and swift data transfer between educational apps is essential to minimizing faculty's cognitive burden-of-discovery and strengthening institution-wide LMS adoption rates. For instance, the enduring utility of Canvas's file, assignment and quiz management apps observed for Virginia Tech faculty is in part because of their easy integration with grading apps. This lets course staff grade assessments without worrying about manual data imports or data corruption. Figure \ref{fig_app_share} describes the frequently-used third-party apps at Virginia Tech, overall and at the Department of Computer Science. For the latter, the commonly used services in these app-suites are discussion forums and course content management (Piazza, Top Hat), exam management (WebCAT), programming instruction and interactive visualizations (OpenDSA, BlockPy, CodeWorkout), etc. Used frequently often by undergraduate courses on programming, algorithms and software engineering, these apps do not affect course GPA and DFW rates (considered together or individually) in the department. While they offer seamless integration with LMS tools for course, student and exam management, a majority of these apps lack one-to-many LTI connections which allow cross-course access, collaboration and research features. Lack of essential interoperability and ubiquitous access features (such as reusing legacy materials in future iterations of the course) often restricts LMS use to housekeeping functions (modest or one-off use of announcements, files, and gradebook), discourages research into new pedagogies, and fuels poor returns on the institutional capital investment into LMS tools.

\subsection{(RQ2) How can platform analytics inform institutional support and management?} 
LMS administrators and instructional designers can use DOU to support departmental resource allocation, faculty development, course evaluation, and LMS evangelism. In this section, we discuss our experiences with these DOU use-cases at Virginia Tech, and the lessons learned.

\subsubsection {Supporting institutional resource allocation and professional development}
DOU can inform the allocation of teaching support at the department or college level. It can serve as a data-driven signal of the need for direct, personalized interventions or additional teaching support for faculty micro-cohorts. For instance, in tables \ref{tab_hyp}, \ref{tab_hyp_one} and \ref{tab_hyp_two}, the hypothesis \textbf{H10} brings the relative utility of a comprehensive professional skills program into question (compared, for instance, to number of TAs in \textbf{H9}), as the cohort is at best indifferent to `deeper' LMS use. A similar picture emerges in table \ref{tab_low_adop} where cohorts with little to no teacher support staff result in a substantial fraction of low DOU courses. The availability of digital skills training does not appear to affect the majority of these courses. Low DOU courses often frequent the cohorts with postgraduate content, online audiences, and low \#TAs, and faculty training alone does not appear effective in alleviating the cognitive burden of discovery required for rapid LMS adoption. We also observe, for instance, that according to hypothesis \textbf{H7} (table \ref{tab_hyp_indiv}), higher average course GPAs and lower DFW rates are linked to higher quiz submission and syllabus DOUs. Undergraduate management, leadership and policy courses make up 35\% of the course cohort with no online quiz submissions, while STEM undergraduate courses make up a majority of high Canvas quiz submissions. Natural resource management (graduate) and physics (undergraduate) courses make up 40\% of courses with no online syllabi. Identifying micro-cohorts with deficiencies in app-level LMS use can help colleges and departments develop online training tools, allocate technical support especially during a pandemic, and improve outreach and faculty buy-in towards new LMS tools. 

\begin{table}[t]
\caption{Low-DOU course frequencies by context}
\label{tab_low_adop}
\begin{tabular}{cccc}
\toprule
\bfseries Context & \bfseries \% & \bfseries Context & \bfseries \% \\
\midrule
Grad $\land$ Low \#TA $\land$ Low enroll & 41 & Grad $\land$ Online $\land$ Low \#TA $\land$ No skills & 51 \\
Low enroll $\land$ Low \#TA $\land$ No app use & 34 & 
Grad $\land$ Online $\land$ Low \#TA & 49\\
Low \#TA $\land$ No app use & 31 & Grad $\land$ Online $\land$ No skills & 41\\
\bottomrule
\end{tabular}
\end{table}

\begin{figure}[t]
\centering
\includegraphics[width=0.55\textwidth]{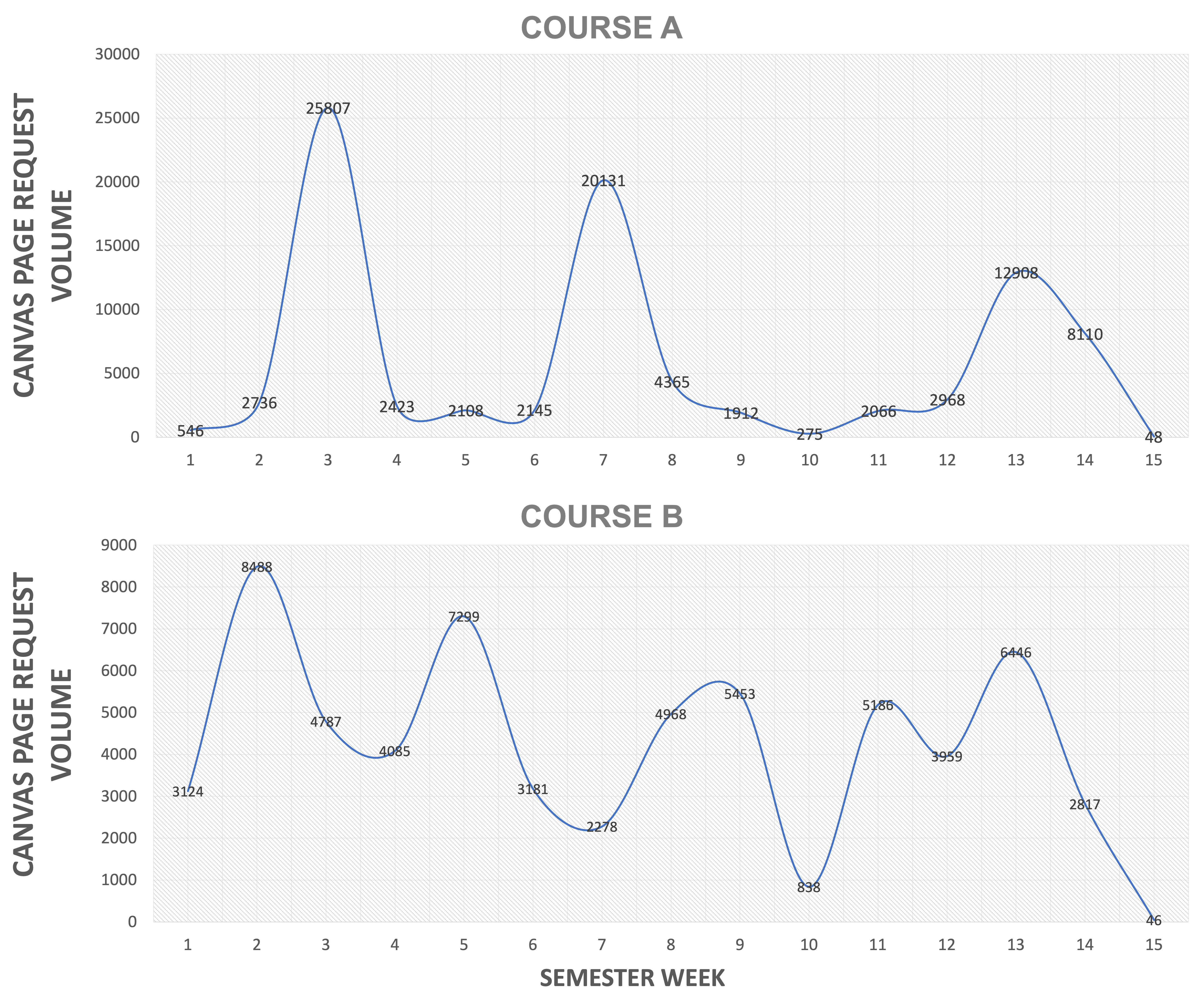}
\caption{LMS page request volume and types for the `junk-drive' use-case. Page requests for course A (top) occur around three key course milestone deadlines, and reflect a variety of LMS resources (backup quiz submissions, show courses, create quizzes, show course activity summary, show quizzes). Course B (bottom) uses the LMS site as a storage drive with no student activity, with nearly all page requests belonging to a single category (show files).}
\label{fig_courseab}
\end{figure}

\subsubsection {Supporting design assessment and LMS evangelism}
DOU can point LMS administrators to faculty preferences about the use of LMS tools and legacy apps, and their broader reasons like trade-off between teaching and research responsibilities, faculty self-efficacy, and cognitive burden of discovery \citep{coates2005critical, west2007understanding}. Table \ref{tab_low_adop} describes some example low-adoption cohorts which highlight the connection between student viewership, course modality and logistics (low DOU courses are 21\% of the dataset overall). In a series of expert reviews at Virginia Tech (N=7, 3 instructional designers, 2 faculty members, 2 software development architects), we identified three distinct types of low-adoption use-contexts, and their implications for design interventions.

\paragraph{\textbf{Junk-drive}}
According to table \ref{tab3}, the overall frequency of low-DOU courses in the dataset is 21\%. Compare these with the frequency of low-adoption courses for several micro-cohorts in table \ref{tab_low_adop}. These frequencies echo the connection between instructor and student engagement and how key aspects of course content and logistics might affect the system and information quality experienced by students while interacting with the LMS. 
An interesting scenario emerges in the connection between viewership and DOU. Figure \ref{fig_courseab} visualizes the weekly average pageviews for two STEM courses with medium weekly viewership and vastly different DOUs. The share of page requests by category (application controller::action) reveals the differences in LMS utilization: course B is primarily being used as a file drive despite having gone through design intervention. Course A, on the other hand, reveals heavy LMS use around two key deadlines for the course and a surge in page views early on in the semester (corresponding to add-drop period for the term). We identified 29 low-DOU high-viewership courses in the spring 2023 dataset. Table \ref{tab_usecases} details their attributes. They are likely to be undergraduate and STEM courses. These courses often have small class sizes; about half of them have teaching assistance and report digital skills training, while none report the use of third-party apps. This micro-cohort is an important example of the potential for continual LMS evangelism and instructional support in order to nudge LMS adoption beyond passive use. 

\begin{table}[t]
\caption{Compositions of the junk-drive, gradebook-only and access-portal course cohorts in spring 2023. For instance, 42\% of the gradebook-only cohort has a large class size, and 80\% has low \#TAs (if any).}
\centering
\label{tab_usecases}
\begin{tabular}{cccc}
\toprule
\bfseries Course attribute & \bfseries Junk-drive (\%) & \bfseries Gradebook-only (\%) & \bfseries Access portal (\%)\\
\midrule
Undergraduate & 79 & 75 & 40\\
STEM & 90 & 73 & 80\\
Online & 17 & 24 & 60\\
3rd-party app use & 0 & 1 & 100\\
Enrollment (Lo, Hi) & 62, 38 & 78, 22 & 80, 20\\
Viewership (Lo, Hi) & 0, 100 & 83, 17 & 100, 0\\
\#TAs (Lo, Hi) & 52, 48 & 71, 29 & 100, 0\\
Skills training & 52 & 56 & 60\\

\bottomrule
\end{tabular}
\end{table}

\paragraph{\textbf{Gradebook-only}}
According to table \ref{tab_usecases}, exclusive use of the Canvas gradebook likely coincides with medium to high-enrollment, undergraduate (75\%), STEM (73\%) courses with an abundance of labs, recitations and group projects. Digital skills training is particularly ineffective for this cohort, which brings to attention its scarce teaching support staff and the likely use of third-party apps unaccounted for in LMS data. It simultaneously points to the need for design interventions that help reduce the cognitive burden of faculty looking to make a fuller transition to LMS discussion forums, groups and assessments, especially at scale.

\paragraph{\textbf{Access-portal}}
This micro-cohort refers to course sites that are collections of links to third-party apps. Per table \ref{tab_usecases}, these courses are often graduate (60\%), STEM (80\%) and unresponsive to digital skills training. Such an extreme reliance on these apps is often a function of both department-level precedents and faculty-perceived ease-of-use. This implies that a design intervention for this micro-cohort should make a particular note of faculty's technology self-efficacy and access to teacher support (note the lack of teaching assistance) in end-of-semester quality assessments.

\begin{table}[t]
\caption{COVID-19 pandemic response planning: overall course DOUs before and after the institution-wide transition to emergency remote teaching at Virginia Tech (spring 2020).}
\centering
\label{tab_pandemic_overall}
\begin{tabular}{p{1.5cm} p{3cm} p{2.8cm} p{2cm}}
\toprule
\bfseries DOU & \bfseries Before transition & \bfseries After transition & \bfseries \% Change\\
\midrule
Low & 3932 & 3667 & -6.49\%\\
Medium & 2365 & 2288 & -3.26\%\\
High & 673 & 1004 & +49.18\%\\

\bottomrule
\end{tabular}
\end{table}

\begin{table}[]
\caption{COVID-19 pandemic response planning: \% change in course DOUs by LMS tool after institution-wide transition to emergency remote teaching at Virginia Tech (spring 2020).}
\centering
\label{tab_pandemic_resource}
\begin{tabular}{cccccccccc}
\toprule
\bfseries DOU & \bfseries An & \bfseries D & \bfseries F & \bfseries S & \bfseries $A_{D}$ & \bfseries $A_{S}$ & \bfseries $Q_{D}$ & \bfseries $Q_{S}$ & \bfseries G\\
\midrule
Low & -7\% & -9\% & -4\% & -1\% & -13\% & -54\% & +20\% & -28\% & -4\%\\

Medium & -23\% & +63\% & - & -0.7\% & -4\% & +12\% & +27\% & - & -\\

High & +44\% & +40\% & +6\% & +2\% & +15\% & +24\% & +30\% & +36\% & 6\%\\

\bottomrule
\end{tabular}
\end{table}

\subsubsection{Supporting pandemic-era transition to remote teaching}

In the spring of 2020, an institution-wide policy of emergency remote teaching was rapidly enacted by Virginia Tech IT leadership, in response to the COVID-19 pandemic. System administrators began with a DOU analysis conducted at the beginning of the term to determine key low-DOU course clusters (upper level STEM and general education coursework), frequent high DOU LMS features (typically the ones with lowest cognitive burden-of-discovery like files and gradebook), and frequent low DOU LMS features (quiz and assignment delivery). The administrators facilitated a rapid transition to remote teaching over a period of two weeks by focusing their support on low-DOU instructors. They designed training classes, in-person consultations, and in-depth documentation focusing on delivery and submission of assignments and quizzes via Canvas. According to table \ref{tab_pandemic_overall}, the IT transition team was able to increase the total number of high DOU courses by over 49\%. Table \ref{tab_pandemic_resource} breaks down the post-COVID DOU gains by LMS tools. Three key takeaways emerge. First, the smallest gains were observed in files and gradebook modules, which hints at an abundance of junk-drive and gradebook-only courses prior to the transition. Instructors new to an LMS tend to first explore the tools they can utilize without significant cognitive effort. Second, the team found that post-transition, instructors' use of announcements (for bulk communication within the LMS) and discussion forums (as a replacement for in-class, face-to-face interaction) increased significantly (+44\%, and +40\%, respectively). Third, the increase in courses with high DOUs for assignment delivery (+15\%), assignment submission (+24\%), quiz delivery (+30\%), and quiz submission (+36\%) is often at the expense of low DOU courses in the same category. This suggests that in favoring online course assessments, Virginia Tech faculty responded to the transition team's focused development and support initiative. While an uptick in overall DOUs is expected during a transition to remote teaching, we contend that if the transition team had not been able to marshall their resources and provide directed support based on DOU analysis, we would observe an abundance of low $\rightarrow$ medium DOU growth. The instances of low $\rightarrow$ high and medium $\rightarrow$ high DOU growth suggest that our pre-transition analyses facilitated an actionable assessment of key faculty needs and a focused adjustment of the IT training and support regimen which maximized its impact.


\subsection{Limitations and threats to validity}
To the best of our assessment, our study is a first institution-wide analysis of LMS use, with over thirty thousand college courses. There are, however, several key limitations of our analysis, owed primarily to (1) the geographic and institutional bounds of the dataset, and (2) the challenging dollar-cost and time overhead of the petitioning, storage, analysis, and compliance requirements of terabyte-scale LMS data. One, our study data is sourced from a single LMS commissioned at Virginia Tech. The focus on a single institution may limit the generalizability of our conclusions beyond Virginia Tech's peer institutions \cite{vtpeers}. Two, the historical preference for the use of older LMSs, for instance by senior faculty, might produce uncertainty in longitudinal DOU measurements for Canvas LMS. The use of older LMSs and comparable LMS software at peer institutions needs to be modeled in future work. Three, our datasets do not capture the fine-grained use of third-party apps. Availability of this data is largely vendor-restricted, so we can only account for apps commissioned through the LMS (refer to section 6 for related future work vectors). Four, our study hypotheses are not exhaustive. We focus on a broad array of course-level metavariables which affect LMS use and potentially impact learning and policy outcomes. Because of sparsity of available data, however, we are unable to account for specialized learning environments and funding inequities at the department and college level. We hope to investigate their relationship with DOU in our future work.

\section{Conclusions and Future Work}
In depth-of-use, we devise a multi-factor, resource-specific view of LMS utilization. DOU helps us examine a variety of use-contexts in faculty adoption of LMS services and assess institutional process effectiveness. Our hypothesis tests reveal that the needs for scale, ubiquitous access and interoperability drive a broad swath of courses across departments towards higher LMS use. We also discover that DOU helps us isolate low-adoption course cohorts, allocate institutional support, and reflect on faculty preferences, technology limitations, and administrative policies that might drive these cohorts. Our research aims to combine expertise from course planning, policy design and quality assurance in order to test multi-level claims of efficacy and recommend interventions that leverage the totality of contextual evidence of historical LMS use.  

Our dataset and analysis describes all Canvas course sites commissioned between, and including, spring 2021 and spring 2023, at Virginia Tech. Its scope can be broadened in several important ways. We examine these as directions of future work as follows. To aid generalizability, we intend to reproduce our analyses for Scholar LMS - in use prior to Canvas - at Virginia Tech. We also plan to compare our results with courses hosted aboard Canvas at peer institutions. Beyond between-LMS and between-institution studies, we hope to hypothesis-test DOU as a function of course modality (flipped and blended classrooms \citep{dias2014towards}), and content and system quality (example pervasiveness \citep{warnick2009imitation}, cognitive task models \citep{masapanta2018systematic, prasad2018developing}, early availability of course content, site aesthetics \citep{martin2008usability}, mobile platform support \citep{casany2012analyzing},  accessibility \citep{wilcox2016canvas} and trust \citep{hassan2019trust, hassan2021learning}), in order to analyze their impact on the usability of LMS services.

In recent years, educators and IT administrators have been widely interested in the use of emerging tools like generative AI \cite{mao2024generative}, virtual and mixed reality \cite{radianti2020systematic}, and short-form video \cite{escamilla2021incorporating} to support higher learning. These technologies present a promising array of use-cases in teaching and learning, such as supporting sensemaking \cite{suh2023sensecape}, productivity \cite{llm-writing}, groupwork \cite{holstein2022designing}, and assessment \cite{llm-review}. We identify the development of DOU taxonomies for these technologies, and validating their relationship with learning outcomes, as crucial vectors of future work. These taxonomies can help instructors evaluate teaching efficacy, and provide IT administrators with decision information on pilot testing, budgeting, licensing, and infrastructure management for new software. We also envision a broader role for LMS-hosted content recommender systems \cite{manouselis2011recommender} as vehicles for faculty outreach, micro-learning, professional development, and personalized technical support. We seek to evolve the DOU measurement and validation strategies in this study to support these emerging technologies at scale. 

We also plan to account for user-activity within third-party apps hosted by the LMS. We plan to collaborate with several app vendors to better understand the relative satisfaction with interactional and content quality these apps might provide. Finally, the scope of our analysis is interpretive in that it examines the observed LMS usage as a function of high-level course meta-characteristics. In our future work, we plan to concurrently model instructor preferences, habits, and values that make up the said usage. We plan to incorporate instructor work experience and familiarity with instructional design practices in our analyses. We also plan to collect qualitative feedback from instructors and students, using semi-structured interviews and online academic forum analyses \citep{hassan2019exploring, hassan2019bias} for key low-adoption micro-cohorts to better summarize and validate these reasons. 

\bibliographystyle{ACM-Reference-Format}
\bibliography{sample-base}

\clearpage

\section*{Appendix}
\setcounter{section}{1}

\appendix
\section{Hypothesis tests by semester}
Tables \ref{tab_hyp_one} and \ref{tab_hyp_two} describe the hypothesis-tests for DOU and course metadata from spring 2021, fall 2021, spring 2022, and fall 2022 academic terms at Virginia Tech. 

\begin{table}[h]
\caption{Spring 2021 and Fall 2021: Hypothesis-testing the relationship between DOU and key course attributes.}
\centering
\label{tab_hyp_one}
\begin{tabular}{lll}
\hline\noalign{\smallskip}
\bfseries Hypothesis & \bfseries (S21) $t$, \bfseries $F$, \bfseries $H$ & \bfseries (F21) $t$, \bfseries $F$, \bfseries $H$\\
\noalign{\smallskip}\hline\noalign{\smallskip}
\textbf{H1}: Undergraduate & 9.4**, 88.6**, 88.7** & 11.5**, 131.2**, 128.6**\\
\textbf{H2}: STEM & -5.2*, 27*, 26* & -5*, 26.8*, 25.7*\\
\textbf{H3}: Online & 8.6**, 74.7**, 68** & 3.9*, 15.2*, 15.8*\\
\textbf{H4}: App use & 3.3*, 11*, 12.5* & 5.5*, 29.7*, 30.2*\\
\textbf{H5}: Enrollment & -, 89.6**, 1e3** & -, 66.6**, 1e3**\\
\textbf{H6}: Viewership & -, 1e3**, 2e3** & -, 809**, 2e3**\\
\textbf{H7}: GPA & -, 81.8**, 187** & -, 24**, 63.5**\\
\textbf{H8}: DFW & -, 31.5**, 130** & -, 12*, 98**\\
\textbf{H9}: \#TA & -, 43**, 488** & -, 31**, 477**\\
\textbf{H10}: Skills & -2*, 4.2*, 4.1* & 1.9*, 3.8*, 4*\\
\noalign{\smallskip}\hline
\multicolumn{3}{l}{*$\alpha=0.05$, stat. signif. $p<=\alpha \land $ $p>1e{-10}$, **$p<1e{-10}$}\\
\end{tabular}
\end{table}

\begin{table}[h]
\caption{Spring 2022 and Fall 2022: Hypothesis-testing the relationship between DOU and key course attributes.}
\centering
\label{tab_hyp_two}
\begin{tabular}{ccc}
\hline\noalign{\smallskip}
\bfseries Hypothesis & \bfseries (S22) $t$, \bfseries $F$, \bfseries $H$ & \bfseries (F22) $t$, \bfseries $F$, \bfseries $H$\\
\noalign{\smallskip}\hline\noalign{\smallskip}
\textbf{H1}: Undergraduate & 9.3**, 87.6**, 87.6** & 11.6**, 135**, 132.8**\\
\textbf{H2}: STEM & -4.1*, 17.2*, 16.8* & -6.6**, 44.5**, 42.8**\\
\textbf{H3}: Online & 2*, 4*, 4.3* & 5*, 25.4*, 25.9*\\
\textbf{H4}: App Use & 6.9**, 47.5**, 47.4** & 7.6**, 58.7**, 59**\\
\textbf{H5}: Enrollment & -, 82**, 1.1e3** & -, 87.7**, 1e3**\\
\textbf{H6}: Viewership & -, 267**, 2e3** & -, 1e3**, 2e3**\\
\textbf{H7}: GPA & -, 35.7**, 106** & -, 31.7**, 93.2**\\
\textbf{H8}: DFW & -, 12.2*, 95.4** & -, 18.2*, 1e2**\\
\textbf{H9}: \#TA & -, 65.5**, 462** & -, 51**, 447**\\
\textbf{H10}: Skills & 0.48, 0.23, 0.2 & 1.3, 1.7, 1.9\\
\noalign{\smallskip}\hline
\multicolumn{3}{l}{*$\alpha=0.05$, stat. signif. $p<=\alpha \land $ $p>1e{-10}$, **$p<1e{-10}$}\\
\end{tabular}
\end{table}

\end{document}